\documentclass[a4paper]{jpconf}
\usepackage{graphicx}
\usepackage{url}
\usepackage{amsmath}
\usepackage{amssymb}
\usepackage{latexsym}
\usepackage{citesort}

\newcommand{\ep}{\varepsilon}
\renewcommand{\sec}{Sec.~}
\newcommand{\fig}{Fig.~}
\newcommand{\eq}{Eq.~}
\newcommand{\eqs}{Eqs.~}
\newcommand{\nr}[1]{(\ref{#1})}
\newcommand{\rx}{x}
\newcommand{\ba}{\begin{eqnarray}}
\newcommand{\ea}{\end{eqnarray}}
\allowdisplaybreaks

\begin{document}

\title{Fun with higher-loop Feynman diagrams}
\author{Thomas Luthe$^1$ and York Schr\"oder$^2$}
\address{$^1$ Institut f\"ur Theoretische Teilchenphysik, Karlsruhe Institute of Technology (KIT), Karlsruhe, Germany}
\address{$^2$ Grupo de Fisica de Altas Energias, Universidad del Bio-Bio, Casilla 447, Chillan, Chile}
\ead{thomas.luthe@kit.edu, yschroeder@ubiobio.cl}

\begin{abstract}
We review recent progress that we have achieved in evaluating the class of fully massive vacuum integrals at five loops. After discussing topics that arise in classification, evaluation and algorithmic codification of this specific set of Feynman integrals, we present some selected new results for their expansions around $4-2\ep$ dimensions.
\end{abstract}

\section{Introduction}

In high-energy physics experiments performed at current colliders such as the LHC, the flood of precision data requires matching theoretical efforts, in order extract the underlying event's structure.
This is particularly relevant for Run II, where important processes are plagued with large QCD backgrounds, requiring high-precision theory calculations.

To this end, we will showcase a few techniques and results related to investigations of the structure of higher-loop Feynman integrals which provide one of the basic building blocks of high-precision perturbative calculations within elementary particle physics. We will discuss new results on the current (five-)loop frontier, which constitute basic building blocks for important quantities such as anomalous dimensions in gauge field theories such as QCD.

\section{Classification}
\label{se:classification}

Let us consider fully massive 5-loop tadpoles, in Euclidean space-time, having the same mass $m$ in all propagators $1/(q_i^2+m^2)$. 
Due to the absence of external legs there are no further scales, such that all integrals scale trivially with the mass (which we can hence set to unity), such that our problem corresponds to computing scale-free functions of the space-time dimension $d$ only. The corresponding 4-loop problem has been solved already more than a decade ago in 4d \cite{Laporta:2002pg} as well as 3d \cite{Schroder:2003kb}.

\begin{figure}
\begin{center}
\includegraphics[width=\textwidth]{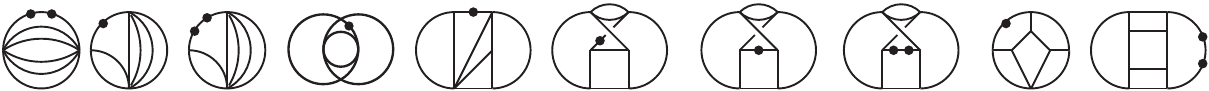}
\end{center}
\caption{\label{fig:dots}Some examples of 5-loop master integrals with 'dots' that we choose as elements of our basis. Each solid line corresponds to a massive propagator with unit power, while each dot represents an extra power. Note that we push all dots on a single line, which is the most natural choice with respect to the difference equation setup for their evaluation.}
\end{figure}

To fully define the 5-loop integral family, we need 15 propagators (lines), for which we choose
\ba
\label{eq:family}
q_i\in\{k_1, k_2, k_3, k_4, k_5, k_{13}, k_{14}, k_{15}, k_{23}, k_{24}, k_{25}, k_{35}, k_{45}, k_{124}, k_{34}\} \;,
\ea
where $k_{a\dots bc}=k_a+\dots+k_b-k_c$.
Trivalent 5-loop graphs have 12 lines, and there are four independent such sectors (one of them corresponding to the last diagram of \fig\ref{fig:dots}).
We label different integral sectors by their binary representation (where a 1 at position $j$ corresponds to a line with momentum $q_j$ in the representation \eq\nr{eq:family}), identify unique graphs, find all isometries and corresponding momentum shifts, and choose the largest representative from each class.

Furthermore, using the freedom of normalization of the integral measure, we find it convenient to divide out a factor of [1-loop tadpole]$^{\mathrm \#loops}$ from each integral, achieving a convention-independent normalization of loop integrals. 
It is useful to recall that in 4d, the [1-loop tadpole]$\sim 1/\ep$, while it is finite in 3d.

To show some 5-loop numerology: of the $2^{15}=32768$ possible sectors that can be constructed from our list of momenta, 5151 do not correspond to a Feynman graph; 1941 of them are trivially zero (in dimensional regularization, since they do not depend on one of the loop momenta), while a further 3625 vanish non-trivially (since they can be shifted such that they become independent of one of the integration momenta); 21962 sectors can be transformed (by linear momentum shifts with unit Jacobian) to a larger representative, while 22 others never occur in a sane Feynman diagram labelling (since they would require a shift with Jacobian 1/2); leaving us with 48 unique irreducible 5-loop master sectors, plus 19 unique master sectors that correspond to products of lower-loop integrals. To compare, at 1/2/3/4-loop there were 1/1/3/10 unique sectors (plus 0/1/2/6 factorized ones).

While each of the 48+19 sectors contains at least one master integral (which can be chosen to be the 'corner integral' having all lines with power one), a (small) IBP reduction reveals that some sectors contain multiple master integrals.
In particular, we need in 62 (+3 factorized ones) additional masters with 'dots', examples of which are shown in \fig\ref{fig:dots}.
Recall that at 1/2/3/4-loop there were only 0/0/0/3 masters with 'dots' \cite{Schroder:2002re}.

\section{Evaluation}

Generalizing our set of master integrals to carry a symbolic power $\rx$ on one of their lines, we can perform an IBP reduction on this modified set of masters and derive a hierarchical system of difference equations for all of them. Symbolically, 
\ba
I(\rx) &\equiv& \int\frac{1}{D_1^{\rx} D_2...D_{N}}
\;,\qquad
\sum_{j=0}^{R} p_{j}(\rx)I(\rx+j)=F(\rx)
\label{eq:deq}
\ea
where the integer $R$ is called order of the equation, the $p_j$ are rational functions (in two variables, $\rx$ and $d$), and the functions $F$ on the right-hand sides contain integrals that are simpler (according to the ordering prescription chosen).
Boundary conditions for these equations are simple, and can be derived by shrinking a line (i.e., at $\rx=0$) or be related to lower-loop integrals (i.e., expanding the integrand for $\rx\gg1$).

Typically, one finally wants to extract the value for $I(1)$. 
To this end, one needs to solve the difference equation; this is possible explicitly for $R=1$, while higher-order equations call for a much more general setup \cite{Laporta:2001dd}.
Laporta's recipe suggests to first attempt a general solution in terms of factorial series, schematically:
\ba
\label{eq:sum}
I(x) &=& I_0(x)+\sum_{j=1}^R I_j(x) \;,\quad\mbox{where}\quad
I_j(x) = \mu_j^x\sum_{s=0}^\infty a_j(s) \,\frac{\Gamma(x+1)}{\Gamma(x+1+s-K_j)} \;.
\ea
The difference equation then determines the special solution $I_0$, fixes the constants $\mu_j$, $K_j(d)$ and implies a recurrence relation for the series coefficients $a_j(s)$ of the homogeneous solutions. 
Second, one needs a boundary condition for fixing, say, $a_j(0)$; it is useful to observe decoupling at large $x$, providing a particularly simple and automatable way to compute the necessary boundary values:
\ba
I(x) &=& \int_{k_1} \frac{g(k_1)}{(k_1^2+1)^x}
\quad\Rightarrow\quad
I(x) \sim (1)^x x^{-d/2} g(0) \;.
\ea
This is then to be compared with the large-$x$ behavior of the factorial series, $\sum_j \mu_j^x x^{K_j} a_j(0)$.

Third, having all analytic ingredients for their solution in terms of infinite sums at hand, to obtain actual results for the master integrals it is most useful to resort to a numeric treatment. This involves truncating the sum in \eq\nr{eq:sum} at some $s_{\rm max}$, evaluating $R$ consecutive values of $I(x)$ around some large $x$ (where the factorial series converges well), and using the difference equation \eq\nr{eq:deq} to push their argument down to the required integrals, such as $I(1)$.

\section{Implementation}

The above method brings with it a number of advantages:
it allows for a high level of automation;
works well with divergent integrals;
does not rely on specific function classes;
gives high-precision results for arbitrary $\ep$ orders;
allows for expansions around any dimension; 
provides simple but highly non-trivial cross-checks by putting $x$ on different lines.

However, we have found that, following the program proposed by Stefano Laporta \cite{Laporta:2001dd}, one is faced with some problems and limitations when treating complex problems: 
the method is of limited use for multi-scale integrals;
the complexity of coefficients in high-order equations can grow enormously;
one typically obtains recurrence relations of high orders;
in numerical evaluation, one often faces instabilities of the factorial series solutions. 

To alleviate these problems and make feasible the computation of our set of 5-loop massive tadpoles, we had to considerably change the traditional setup.
Much of our progress can be attributed to a number of fixes done to the original proposal. To name a few key ingredients \cite{TLdiss}: 
we use coupled IBP equations, in order to tame the growth of complexity; 
we reduce (the orders of) recurrence relations, essentially by re-using the linear solver developed for the system of difference equations;
our code predicts instability factors, in order to assign sufficient numerical precision.

\section{Results}

\begin{figure}
\begin{center}
\includegraphics[width=\textwidth]{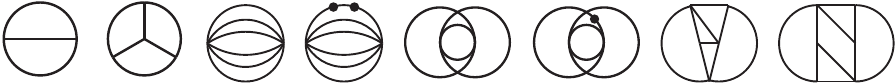}
\end{center}
\caption{\label{fig:samples}Sample integrals for which results are given in \eqs\nr{eq:7}, \nr{eq:63}, (\ref{eq:28686}-\ref{eq:31420}).}
\end{figure}

In order to be specific, let us focus on results in $d=4-2\ep$ here, recalling that we normalize all our integrals with the corresponding power of the massive 1-loop tadpole.
Given our high-precision numerical results, one might wonder whether it is possible to determine some of the coefficients in analytic form. 
To this end, we employ the integer-relation finder \texttt{PSLQ} \cite{MR1489971} together with an educated guess of a basis of irrationals (such as e.g.\ the well-classified set of multiple zeta values and/or alternating Euler sums), to guess some of these coefficients. 
It helps to keep the basis as compact as possible, so in order to absorb single powers of $\pi$ as well as powers of $\ln3$, we define
\begin{eqnarray}
h_n &\equiv& \sum_{k=0}^\infty\frac{\Gamma(k+1/2)}{\Gamma(k+1)\Gamma(1/2)}\,\frac{(3/4)^k}{(2k+1)^n}
= {}_{n+1}F_n[\underbrace{\{1/2,..,1/2\}}_{n+1},\underbrace{\{3/2,..,3/2\}}_n,3/4] \;,
\\
H_n &\equiv& h_n+h_1\,\mbox{Coefficient}\Big[1-\frac{3^{\ep/2}\Gamma(1-\ep)}{\Gamma^2(1-\ep/2)}+{\cal O}(\ep^n),\ep,n-1\Big]\;,\\
&\Rightarrow& 
H_1 = h_1 = \frac{2\pi}{3\sqrt3}\;,\quad
H_2 = h_2-\frac12h_1\ln3\;,\quad
H_3 = h_3-\frac{h_1}{8}(\ln^23+2\zeta_2)\;,\quad 
etc.
\end{eqnarray}
The specific combinations $H_n$ have actually been inspired by the 2-loop sunset (leftmost diagram of \fig\ref{fig:samples}), which can be written to all orders in terms of the $H_n$, the first few of which read\footnote{Regarding our notation for integral labels: $I_{a,b,c}$ refers to the master integral with binary code $a$ as explained in \sec\ref{se:classification}, having propagators with unit powers, except line number $b$, which carries exponent $c$ (or $c-1$ 'dots').}
\begin{eqnarray}
\label{eq:7}
I_{7.1.1} = -\frac{3}{2}-\frac{3}{2}\,\ep+(9 H_2-3)\,\ep^{2}+(9 H_2-18 H_3-6)\,\ep^{3}+(18 H_2-18 H_3+36 H_4-12)\,\ep^{4}\;.
\end{eqnarray}
Furthermore, in order to absorb powers of $\ln2$, let us define as elements of the MZV basis
\begin{eqnarray}
a_n &\equiv& \sum_{k=0}^\infty\frac{1}{2^k\,k^n} = \mbox{PolyLog}[n,1/2]={\rm Li}_n(\frac12)\;,\\
A_n &\equiv& a_n+(-1)^n\frac{\ln^n2}{n!}\Big(1-\frac{n(n-1)}{2}\,\frac{\zeta_2}{\ln^22}\Big)\;,\\
&\Rightarrow&
A_4 = a_4-\zeta_2\frac{\ln^22}{2\cdot2!}+\frac{\ln^42}{4!}\;,\quad
A_5 = a_5+\zeta_2\frac{\ln^32}{2\cdot3!}-\frac{\ln^52}{5!}\;,\quad 
etc.,
\end{eqnarray}
such that e.g.\ the 3-loop mercedes-type integral (second diagram of \fig\ref{fig:samples}) reads
\begin{eqnarray}
\label{eq:63}
I_{63.1.1} = +(0)\,\ep^{0}+(0)\,\ep^{1}+(-2 \zeta _3)\,\ep^{2}+(-16 A_4+27 H_2^2+\frac{34 \zeta _2^2}{5})\,\ep^{3}+\dots\;.
\end{eqnarray}

Our numerical approach has passed a number of cross-checks with flying colors: placing the power $x$ of each generalized master on topologically inequivalent lines, the values at $x=1$ coincide; all known results up to 4-loop level \cite{Laporta:2002pg,Czakon:2004bu,Schroder:2005va,Chetyrkin:2006dh} have been reproduced or improved; all factorized sectors perfectly agree with the respective products of lower-loop integrals; and the few previously known 5-loop coefficients \cite{Groote:2005ay,Broadhurst:1985vq,Brown:2012ia} have been reproduced.
Let us finally show a couple of new results on the five-loop level (for the rightmost 6 graphs shown in \fig\ref{fig:samples}), using our \texttt{PSLQ} fits for low $\ep$\/-orders (the divergent ones), and truncating the constant term to 50 numerical digits for readability:
\begin{eqnarray}
\label{eq:28686}
I_{28686.1.1} &=& +(-3)\,\ep^{0}+(-\frac{3}{2})\,\ep^{1}+(\frac{13}{24})\,\ep^{2}+(-\frac{1267}{1440})\,\ep^{3}+(-\frac{4193}{3456})\,\ep^{4}+
\nonumber\\&&+135.95072868792871461956492733702218574897992953584\,\ep^{5}+\dots\;,
\\ I_{28686.1.3} &=& +(0)\,\ep^{0}+(\frac{3}{2})\,\ep^{1}+(-\frac{1}{2})\,\ep^{2}+(-\frac{443}{360})\,\ep^{3}+(\frac{95}{216})\,\ep^{4}+
\nonumber\\&&-38.292059175062436961881799538284449799148385376441\,\ep^{5}+\dots\;,
\\ I_{30862.1.1} &=& +(-\frac{3}{5})\,\ep^{0}+(-\frac{27}{10})\,\ep^{1}+(-\frac{4 \zeta _3}{5}-\frac{421}{60})\,\ep^{2}+(-\frac{12 \zeta _2^2}{25}+\frac{24 \zeta _3}{5}+\frac{211}{24})\,\ep^{3}
\nonumber\\&&+(\frac{72 \zeta _2^2}{25}-98 \zeta _3+\frac{32 \zeta _5}{5}+\frac{12959}{48})\,\ep^{4}+
\nonumber\\&&+1143.1838307558764599466030303839590323268318605888\,\ep^{5}+\dots\;,
\\ I_{30862.1.2} &=& +(\frac{1}{5})\,\ep^{0}+(\frac{11}{30})\,\ep^{1}+(-\frac{1}{30})\,\ep^{2}+(-\frac{12 \zeta _3}{5}-9)\,\ep^{3}+(-\frac{36 \zeta _2^2}{25}+\frac{548 \zeta _3}{15}-\frac{1229}{15})\,\ep^{4}+
\nonumber\\&&-102.42854342605587086319606311891941160276036031953\,\ep^{5}+\dots\;,
\\ I_{30231.1.1} &=& +(0)\,\ep^{0}+(0)\,\ep^{1}+(\frac{3 \zeta _3}{5})\,\ep^{2}+(\frac{9 \zeta _2^2}{25}+\frac{21 \zeta _3}{5}+3 \zeta _5)\,\ep^{3}
\nonumber\\&&+(-36 H_2 \zeta _3+\frac{12 \zeta _2^3}{7}+\frac{63 \zeta _2^2}{25}-\frac{21 \zeta _3^2}{5}+27 \zeta _3-\frac{24 \zeta _5}{5})\,\ep^{4}+
\nonumber\\&&-531.32391547725635267943444561495368318398901378435\,\ep^{5}+\dots\;,
\\ I_{31420.1.1} &=& +(0)\,\ep^{0}+(0)\,\ep^{1}+(0)\,\ep^{2}+(0)\,\ep^{3}+(-\frac{36 \zeta _3^2}{5})\,\ep^{4}+
\nonumber\\&&+167.81535305918474061962120601112466233675898298296\,\ep^{5}+\dots\;.
\label{eq:31420}
\end{eqnarray}

\section{Conclusions}

We have studied the class of fully massive vacuum diagrams, which are essential building blocks for a number of phenomenological applications such as QCD thermodynamics, anomalous dimensions, or moments. A classification of all 5-loop integrals contained in this class reveals 48 independent sectors, with 48+62 master integrals (the latter having 'dots').

For the evaluation, we have chosen to employ a hierarchical system of difference equations and factorial series, amenable to automated numerical treatment. The setup required for deriving the difference equations, the recursion relations for factorial series coefficients, as well as the numerical solution has been implemented in the \texttt{C++} code \texttt{TIDE} \cite{TLdiss}, essential parts of which are parallelized. We use \texttt{Fermat} \cite{fermat} for polynomial algebra in 2 variables, and manage to employ a substantial fine-tuning of the Laporta approach, without which these results could not have been derived on the computer hardware that we have at our disposal. At 5 loops, our code had to deal with difference equations of up to order 20, and recurrence relations of up to order 28. 

To date, we have numerical results for 44 of the 48 non-factorized 5-loop master-sectors (including the 'dotted' integrals of those sectors), with about 300 digits accuracy, for at least 10 $\ep$\/-orders around $d=4-2\ep$ and $d=3-2\ep$. These results include all divergent integrals, bringing 5-loop anomalous dimensions within reach (once the integral reduction has been performed).

\ack

The work of T.L.\ has been supported in parts by DFG grants GRK 881 and SCHR 993/2, and he wishes to thank the Bielefeld Graduate School in Theoretical Sciences for facilitating finalizing this project via a BGTS mobility grant.
Y.S.\ acknowledges support from DFG grant SCHR 993/1, FONDECYT project 1151281 and UBB project GI-152609/VC.

\section*{References}
\bibliography{acat16procYS}

\end{document}